\documentclass[tightenlines,superscriptaddress,twocolumn,preprintnumbers,amssymb,nofootinbib]{revtex4-1}
\usepackage{amsmath, amsthm, amssymb} 

\usepackage{graphics,bm}
\usepackage{epsfig}
\usepackage{graphicx}
\usepackage{amsmath}
\usepackage{wrapfig}
\usepackage{color}
\usepackage{tikz}
\usetikzlibrary{arrows,shapes}
\usetikzlibrary{trees}
\usetikzlibrary{matrix,arrows} 			
\usetikzlibrary{positioning}				
\usetikzlibrary{calc,through}				
\usepackage{pgffor}							

\usetikzlibrary{decorations.pathmorphing}	
\usetikzlibrary{decorations.markings}
\tikzset{
    sigmaCT/.style={draw=black, postaction={decorate},
        decoration={markings,mark=at position .99 with {\arrow[draw=black]{>}},mark=at 		 position .99 with {\arrow[
draw=black]{<}}}},
    pionCT/.style={dashed,draw=black, postaction={decorate},
        decoration={markings,mark=at position .99 with {\arrow[draw=black]{>}},mark=at position .99 with {\arrow[draw=black]{<}}}},    
    fermionCT/.style={draw=black, postaction={decorate},
        decoration={markings,mark=at position .5 with {\arrow[draw=black]{>}},mark=at position .99 with {\arrow[draw=black]{>}},mark=at position .99 with {\arrow[draw=black]{<}}}},    
    fermion/.style={draw=black, postaction={decorate},
        decoration={markings,mark=at position .55 with {\arrow[draw=black]{>}}}},
    fermionbar/.style={draw=black, postaction={decorate},
        decoration={markings,mark=at position .55 with {\arrow[draw=black]{<}}}},
    pion/.style={dashed,draw=black, postaction={decorate}},
    sigma/.style={draw=black, postaction={decorate}}
}

\newcommand{\beq}{\begin{equation}}
\newcommand{\eeq}{\end{equation}}
\newcommand{\bqa}{\begin{eqnarray}}
\newcommand{\eqa}{\end{eqnarray}}

\newcommand{\ms}{\overline{\text{\tiny MS}}}

\def\square{\vcenter{\vbox{\hrule height.4pt
          \hbox{\vrule width.4pt height4pt
          \kern4pt\vrule width.3pt}\hrule height.4pt}}}

\begin{document}

\title{Pion condensation and phase diagram 
in the Polyakov-loop quark-meson model}

\author{Prabal Adhikari}
\email{adhika1@stolaf.edu}
\affiliation{St. Olaf College, Physics Department, 1520 St. Olaf Avenue,
Northfield, MN 55057, USA}
\author{Jens O. Andersen}
\email{andersen@tf.phys.ntnu.no}
\affiliation{Department of Physics, Faculty of Natural Sciences, NTNU, 
Norwegian University of Science and Technology, H{\o}gskoleringen 5,
N-7491 Trondheim, Norway}
\author{Patrick Kneschke}
\email{patrick.kneschke@uis.no}
\affiliation{Faculty of Science and Technology, University of Stavanger,
N-4036 Stavanger, Norway}
\date{\today}

\begin{abstract}
We use the Polyakov-loop extended two-flavor quark-meson model as 
a low-energy effective model for QCD to
study the phase diagram in the $\mu_I$--$T$ plane where $\mu_I$
is the isospin chemical potential. In particular, we focus on 
the Bose condensation of charged pions. At $T=0$, the onset of pion condensation
is at $\mu_I={1\over2}m_{\pi}$ in accordance with exact results. 
The phase transition to a Bose-condensed phase is of second order
for all values of $\mu_I$  and in the $O(2)$ universality class.
The chiral critical line joins the critical line
for pion condensation at a  point whose position depends on the
Polyakov-loop potential and the sigma mass.
For larger values of $\mu_I$ these curves are on top of each other.
The deconfinement line enters smoothly the phase with
the broken $O(2)$ symmetry.
We compare our results with recent lattice simulations and find overall
good agreement.

\end{abstract}

\maketitle

\section{Introduction}
The phases of QCD as functions of the baryon chemical  potential
$\mu_B$ or  the  quark  chemical  potential $\mu={1\over3}\mu_B$,
and temperature $T$
have  been  studied  in  detail  since  the  first
phase  diagram  was  proposed  more than fourty years ago 
\cite{raja,alford,fukurev}. 
At  vanishing baryon chemical potential, it is possible to 
perform lattice simulations to calculate the
thermodynamic  functions  and  the  transition  temperature  
associated  with  chiral  symmetry  restoration  and
deconfinement.  For physical quark masses and two flavors,  
the  transition  is  a  crossover  at  a  temperature  of
approximately 155 MeV \cite{lat1,lat2,lat3,lat4}.

At nonzero baryon chemical potential, however, Monte Carlo simulations
are hampered by the so-called sign problem, namely that the
fermion determinant becomes complex. Being complex, 
the usual interpretation of it as part  of a probability 
distribution can no longer be upheld.
The sign problem in QCD at finite baryon density has spurred the interest
in QCD-like theories free of this problem. This includes 
QCD with quarks in the adjoint 
representation \cite{qcdlike1}, two-color QCD \cite{qcdlike2}, 
QCD at finite isospin $\mu_I$ \cite{son}, and
QCD in a magnetic field $B$ \cite{bali}. 
These theories are all interesting in their own right;
QCD at finite isospin and QCD in a magnetic
field are also relevant
for compact stars. In addition, the application of Monte-Carlo methods
allows a direct test of various model approaches in the cases mentioned above.
Such a confrontation of model calculations with lattice simulations
of QCD in a magnetic field has been very fruitful in understanding
their strengths and limitations \cite{dimi,rw}.

Lattice simulations of QCD at finite isospin have been performed
in e.g. Refs. \cite{kogut1,kogut2,gergy1,gergy2,gergy3} with particular
emphasis on Bose condensation of charged pions 
for isospin chemical potentials above the 
zero-temperature critical value $\mu_I^c={1\over2}m_{\pi}$.
Chiral perturbation theory (ChPT) \cite{son,kim,loewe,fragaiso,carigchpt}, 
which is a model-independent low-energy theory for QCD
valid at low densities has been used to study pion condensation.
ChPT predicts a second-order transition, which is in agreement
with lattice simulations.
There have also been a number of other approaches and model 
calculations studying various
aspects of the QCD phase diagram at finite isospin density, 
including the resonance gas model \cite{restoublan}, 
random matrix models \cite{random}, the
Nambu-Jona-Lasinio
(NJL) model \cite{2fbuballa,toublannjl,bar2f,he2f,heman2,heman,ebert1,ebert2,sun,lars,2fabuki,heman3,he3f}, 
the quark-meson (QM) model \cite{lorenz,ueda,qmstiele,patrick}
\footnote{Or their Polyakov-loop extended versions (PNJL and PQM).},
and
effective theory at asymptotically high isospin~\cite{isohigh}.

Finally, we mention that one expects another phase transition at
large isospin chemical potential. In perturbation theory, one-gluon exchange 
gives rise to an effective attractive interaction between $u$
and $\bar{d}$ quarks leading to the formation of Cooper pairs \cite{son}.
The transition from a Bose-Einstein condensate (BEC) 
to a Bardeen-Cooper-Schrieffer (BCS) state is expected
to be an analytical crossover as the symmetry-breaking pattern is the same.

As pointed out in Ref. \cite{lorenz}, there is a mapping of the 
quark-meson model at finite isospin and the corresponding two-color
quark-meson-diquark model at finite baryon chemical 
potential.
The neutral pion $\pi_0$ is replaced by
an isovector triplet ${\boldsymbol \pi}$.
The charged pions $\pi^{\pm}$ are replaced by a diquark-antidiquark pair
$\Delta$ and $\Delta^*$, which instead of being coupled to $\mu_I$
is now coupled to a baryon chemical potential 
$\mu_B$.\footnote{The diquarks are the baryons of two-color QCD.}
Since the gauge groups $SU(2)$ and $SU(3)$ are fundamentally different,
this mapping is valid for the matter sector; once we couple the
QM model to the Polyakov loop, this identification is lost. 

In the present paper, we study the QCD phase diagram at finite temperature
and isospin density using the PQM model.
The main conclusions of our work are
\begin{enumerate}
\item The second order transition to a BEC state. The transition
is in the $O(2)$ universality class.
At $T=0$, the transition
is exactly at $\mu_I={1\over2}m_{\pi}$.
\item The BEC and chiral transition lines meet at a 
point $(\mu_I^{\rm meet},T^{\rm meet})$ 
and coincide for larger isospin chemical potentials $\mu_I$.
\item The deconfinement and chiral transition lines
coincide in the non-condensed phase for a logarithmic Polyakov-loop
potential and a sufficiently low sigma mass.
\item The deconfinement line penetrates smoothly
into the symmetry-broken phase.
\end{enumerate}
These results are in agreement with the recent lattice simulations of
Refs. \cite{gergy1,gergy2,gergy3}.

The paper is organized as follows. 
In Sec. II, we briefly discuss the quark-meson model and in Sec. III we 
calculate the effective potential in the mean-field approximation.
In Sec. IV, we discuss the coupling to the Polyakov loop, while 
in Sec. V, we present the phase diagram in the $\mu_I$--$T$ plane
and compare it to recent lattice results.
In Appendix A, we list a few integrals needed in the calculations, while
Appendix B provides the reader with some details of how the parameters
of the quark-meson model are determined.

\section{Quark-meson model}
The Lagrangian of the two-flavor quark-meson model 
in Minkowski space is 
\bqa\nonumber
{\cal L}&=&
{1\over2}\left[(\partial_{\mu}\sigma)(\partial^{\mu}\sigma)
+(\partial_{\mu} \pi_3)(\partial^{\mu} \pi_3)
\right]
\\&&\nonumber
+(\partial_{\mu}+2i\mu_I\delta_{\mu}^0)\pi^+(\partial^{\mu}-2i\mu_I\delta_{0}^{\mu})
\pi^-
\\&&\nonumber-{1\over2}m^2(\sigma^2+\pi_3^2+2\pi^+\pi^-)
-{\lambda\over24}(\sigma^2+\pi_3^2+2\pi^+\pi^-)^2
\\ && 
+h\sigma+\bar{\psi}\left[
i/\!\!\!\partial
+\mu_f
\gamma^0
-g(\sigma+i\gamma^5{\boldsymbol\tau}\cdot{\boldsymbol\pi})\right]\psi\;,
\label{lag}
\eqa
where $\psi$ is 
a color $N_c$-plet, a four-component Dirac spinor as well as a flavor doublet 
\bqa
\psi&=&
\left(
\begin{array}{c}
u\\
d
\end{array}\right)\;,
\eqa
and $\mu_f={\rm diag}(\mu_u,\mu_d)$,
where $\mu_u$ and $\mu_d$, are the quark chemical potentials,
$\mu_I$ is the isospin chemical potential,
$\tau_i$ ($i=1,2,3$) are the Pauli matrices in flavor space, 
${\boldsymbol\pi}=(\pi_{1},\pi_{2},\pi_{3})$, and
$\pi^{\pm}={1\over\sqrt{2}}(\pi_1\pm i\pi_2)$.

Apart from the global $SU(N_c)$ symmetry, 
the Lagrangian~(\ref{lag}) 
has a 
$U(1)_B\times SU(2)_L\times SU(2)_R$ symmetry for 
$h=0$ and a $U(1)_B\times SU(2)_V$ symmetry
for $h\neq0$. 
When $\mu_u\neq\mu_d$, this symmetry is reduced to 
$U(1)_B\times U_{I_3L}(1)\times U_{I_3R}(1)$ for $h=0$ and
$U(1)_B\times U_{I_3}(1)$ for $h\neq0$. 

The number density associated with a chemical potential $\mu_A$ is
\bqa
n_A&=&-{\partial V\over\partial\mu_A}\;,
\label{na}
\eqa
where $V$ is the effective potential.
The baryon and isospin densities can be expressed in terms of the
quark densities $n_u$ and $n_d$ as
\bqa
\label{nb}
n_B&=&{1\over3}(n_u+n_d)\;,
\\
n_I&=&n_u-n_d\;.
\label{ni}
\eqa
Eqs. (\ref{nb})--(\ref{ni}) together with the chain rule
can be used to derive relations among the
baryon and isospin chemical potentials and the quark chemical potentials.
We have
\bqa\nonumber
n_I&=&-{\partial V\over\partial\mu_I}
\\ \nonumber
&=&-\left({\partial V\over\partial\mu_u}-{\partial V\over\partial\mu_d}\right)
\\
&=&-\left({\partial\mu_u\over\partial\mu_I}{\partial V\over\partial\mu_u}
+{\partial\mu_d\over\partial\mu_I}{\partial V\over\partial\mu_d}\right)
\;.
\eqa
This yields
\bqa
{\partial\mu_u\over\partial\mu_I}
=-{\partial\mu_d\over\partial\mu_I}=1\;.
\eqa
Similarly, we find ${\partial\mu_u\over\partial\mu_B}={\partial\mu_d\over\partial\mu_B}={1\over3}$.
From this, we find the following relations among the chemical potentials
\bqa
\label{rr}
\mu_u&=&{1\over3}\mu_B+\mu_I\;,\\
\mu_d&=&{1\over3}\mu_B-\mu_I\;.
\label{rrr}
\eqa
Introducing the quark chemical potential $\mu={1\over3}\mu_B$ and 
inverting the
relations (\ref{rr})--(\ref{rrr}), we find
\bqa
\mu&=&{1\over2}(\mu_u+\mu_d)\;,\\
\mu_I&=&{1\over2}(\mu_u-\mu_d)\;.
\eqa

\section{Effective potential}
The expectation values of the fields are written as
\bqa
\sigma&=&\phi_0\;,\hspace{0.5cm}
\pi_1=\pi_0
\;,
\eqa
where $\phi_0$ and $\pi_0$ are constant in space. The 
former is the usual chiral condensate, while the 
latter represents
a homogeneous pion condensate.
A pion condensate breaks the $U_{I_3L}(1)\times U_{I_3R}(1)$
symmetry to $U_{I_3V}(1)$ or the $U_{I_3}(1)$ symmetry.
Introducing 
$\Delta=g\phi_0$
and $\rho=g\pi_0$, 
the tree-level potential in Euclidean space
can be written as 
\bqa\nonumber
V_0&=&{1\over2}{m^2\over g^2}\Delta^2
+{1\over2}{m^2-4\mu_I^2\over g^2}\rho^2
\\ &&
+{\lambda\over24g^4}\left(\Delta^2+\rho^2\right)^2
\label{v0}
-{h\over g}\Delta\;.
\eqa
Expressing the parameters in Eq.~(\ref{lag}) terms of 
the sigma mass $m_{\sigma}$, the pion mass $m_{\pi}$, 
the pion decay constant $f_{\pi}$, and the constituent quark mass $m_q$ , we find
\bqa
\label{rel1}
m^2&=&-{1\over2}\left(m_{\sigma}^2-3m_{\pi}^2\right)\;,\,
\lambda=3{\left(m_{\sigma}^2-m_{\pi}^2\right)\over f_{\pi}^2}\;,
\\
g^2&=&{m_q^2\over f_{\pi}^2}\;,
\hspace{2.1cm}
h=m_{\pi}^2f_{\pi}\;.
\label{rel2}
\eqa
Inserting these relations, we can write the tree-level potential as
\bqa\nonumber
V_0&=&
-{1\over4}f_{\pi}^2(m_{\sigma}^2-3m_{\pi}^2){\Delta^2+\rho^2\over m_q^2}
-
2\mu_I^2f_{\pi}^2{\rho^2\over m_q^2}
\\ 
&&+{1\over8}f_{\pi}^2(m_{\sigma}^2-m_{\pi}^2){(\Delta^2+\rho^2)^2\over m_q^4}
-m_{\pi}^2f_{\pi}^2{\Delta\over m_q}\;.
\label{treepot}
\eqa
The quark energies can be read off from 
the zeros of the determinant of the
Dirac operator. One finds 
\bqa
\label{eu}
E_u&=&E(-\mu_I)\;,
\hspace{1cm}
E_d=E(\mu_I)\;,
\\
E_{\bar{u}}&=&E(\mu_I)\;,
\hspace{1.25cm}E_{\bar{d}}=E(-\mu_I)\;.
\label{ed}
\eqa
where we have defined
\bqa
E(\mu_I)&=&
\left[
\left(\sqrt{p^2+\Delta^2}+{\mu_I}\right)^2
+\rho^2\right]^{1\over2}\;.
\eqa
Note that the quark energies explicitly depend on $\mu_I$.
In the following we choose $\mu_I>0$, but similar results are obtained
for $\mu_I<0$.

The one-loop  contribution to the effective potential 
at $T=\mu_B=0$ is
\bqa
V_1&=&-
N_c\int_p\left(E_u+E_d
+E_{\bar{u}}+E_{\bar{d}}\right)\;,
\label{v1}
\eqa
where
the integral is in $d=3-2\epsilon$ dimensions (See Appendix A).
The integral in Eq. (\ref{v1}) 
is ultraviolet divergent and in order to 
isolate the divergences, we need to expand the energies in powers of
$\mu_I$ to the appropriate order. This yields
\begin{widetext}
\bqa\nonumber
V_{\rm div}&=&
-4N_c\int_p\left[
\sqrt{p^2+\Delta^2+\rho^2}
+{\mu_I^2\rho^2\over2({p^2+\Delta^2+\rho^2)^{3\over2}}}\right]
={4N_c\over(4\pi)^2}
\left({e^{\gamma_E}\Lambda^2\over\Delta^2+\rho^2}\right)^{\epsilon}
\left[
\left(\Delta^2+\rho^2\right)^2
\Gamma(-2+\epsilon)
-2\mu_I^2\rho^2\Gamma(\epsilon)
\right]\;.
\\ &&
\eqa
The remainder $V_{\rm fin}$ is finite and reads
\bqa
V_{\rm fin}&=&V_1-V_{\rm div}\;.
\eqa
Note that $V_{\rm fin}$ can be evaluated directly in $d=3$ dimensions.
In the present case, $V_{\rm fin}$ must be evaluated numerically.
Using the expressions for the integrals listed in Appendix A, we can
write the unrenormalized one-loop effective potential
$V=V_0+V_1$ as
\bqa\nonumber
V&=&{1\over2}{m^2\over g^2}\Delta^2
+{1\over2}{m^2-4\mu_I^2\over g^2}\rho^2
+{\lambda\over24g^4}(\Delta^2+\rho^2)^2
-{h\over g}\Delta
\\ 
&&+{2N_c\over(4\pi)^2}\left(\Lambda^2\over\Delta^2+\rho^2\right)^{\epsilon}
\left[\left(\Delta^2+\rho^2\right)^2\left({1\over\epsilon}+{3\over2}\right)
-4\mu_I^2\rho^2{1\over\epsilon}
\right]+V_{\rm fin}
+{\cal O}(\epsilon)
\;,
\eqa
which contains poles in $\epsilon$. These poles are removed by 
mass and coupling constant renormalization. 
In the $\overline{\rm MS}$ scheme this is achieved by 
making the substitutions $m^2\rightarrow Z_{m^2}m^2$,
$\lambda\rightarrow Z_{\lambda}\lambda$,
$g^2\rightarrow Z_{g^2}g^2$, and 
$h\rightarrow Z_{h}h$, where 
\bqa
Z_{m^2}=1+{4N_cg^2\over(4\pi)^2\epsilon}\;,
\hspace{0.4cm}
Z_{\lambda}=1+{8N_c\over(4\pi)^2\epsilon}\left[
g^2-6{g^4\over\lambda}\right]\;,
\hspace{0.4cm}
Z_{g^2}=1+{4N_cg^2\over(4\pi)^2\epsilon}\;,
\hspace{0.4cm}
Z_{h}=1+{2N_cg^2\over(4\pi)^2\epsilon}\;,
\hspace{0.4cm}
\label{zg}
\eqa
The renormalized one-loop effective potential then reads
\bqa\nonumber
V_{\rm 1-loop}&=&
{1\over2}{m^2_{\ms}\over g^2_{\ms}}\Delta^2
+{1\over2}{m^2_{\ms}-4\mu_I^2\over g^2_{\ms}}\rho^2
+{\lambda_{\ms}\over24g_{\ms}^4}\left(\Delta^2+\rho^2\right)^2
-{h_{\ms}\over g_{\ms}}\Delta
\\ 
&&
+{2N_c\over(4\pi)^2}\Bigg\{
\left[\left(\Delta^2+\rho^2\right)^2-4\mu_I^2\rho^2\right]
\log\left({\Lambda^2\over\Delta^2+\rho^2}\right)
+{3\over2}\left(\Delta^2+\rho^2\right)^2
\Bigg\}+V_{\rm fin}\;,
\label{veff}
\eqa
where the subscript ${\overline{\rm MS}}$ indicates that the
parameters are running with the renormalization scale $\Lambda$.
Using $Z_{g^2}$ in Eq. (\ref{zg}) and the wavefunction renormalization factor
$Z_{\phi}=1-{4N_cg^2\over(4\pi)^2\epsilon}$, it is seen that the fields
$\Delta$ and $\rho$ do not run.
In Appendix B, we discuss how one can express the
parameters in the $\overline{\rm MS}$ scheme in terms
of physical masses and couplings. Using 
Eqs. (\ref{sol1})--(\ref{sol5}), the final expression for the one-loop
effective potential in the large-$N_c$ limit becomes
\bqa\nonumber
V_{\rm 1-loop}&=&
\dfrac{3}{4}m_\pi^2 f_\pi^2
\left\{1-\dfrac{4 m_q^2N_c}{(4\pi)^2f_\pi^2}m_\pi^2F^{\prime}(m_\pi^2)
\right\}\dfrac{\Delta^2+\rho^2}{m_q^2}
\\ \nonumber &&
 -\dfrac{1}{4}m_\sigma^2 f_\pi^2
\left\{
1 +\dfrac{4 m_q^2N_c}{(4\pi)^2f_\pi^2}
\left[ \left(1-\mbox{$4m_q^2\over m_\sigma^2$}
\right)F(m_\sigma^2)
 +\dfrac{4m_q^2}{m_\sigma^2}
-F(m_\pi^2)-m_\pi^2F^{\prime}(m_\pi^2)
\right]\right\}\dfrac{\Delta^2+\rho^2}{m_q^2} 
\\ \nonumber &&
-2\mu_I^2f_\pi^2
\left\{1-\dfrac{4 m_q^2N_c}{(4\pi)^2f_\pi^2}
\left[\log\mbox{$\Delta^2+\rho^2\over m_q^2$}
+F(m_\pi^2)+m_\pi^2F^{\prime}(m_\pi^2)\right]
\right\}{\rho^2\over m_q^2}
\\ \nonumber
 & & + \dfrac{1}{8}m_\sigma^2 f_\pi^2
\left\{ 1 -\dfrac{4 m_q^2  N_c}{(4\pi)^2f_\pi^2}\left[
\dfrac{4m_q^2}{m_\sigma^2}
\left( 
\log\mbox{$\Delta^2+\rho^2\over m_q^2$}
-\mbox{$3\over2$}
\right) -\left( 1 -\mbox{$4m_q^2\over m_\sigma^2$}\right)F(m_\sigma^2)
+F(m_\pi^2)+m_\pi^2F^{\prime}(m_\pi^2)\right]
 \right\}\dfrac{(\Delta^2+\rho^2)^2}{m_q^4}
\\ &&
- \dfrac{1}{8}m_\pi^2 f_\pi^2
\left[1-\dfrac{4 m_q^2N_c}{(4\pi)^2f_\pi^2}m_\pi^2F^{\prime}(m_\pi^2)\right]
\dfrac{(\Delta^2+\rho^2)^2}{m_q^4}
-m_\pi^2f_\pi^2\left[
1-\dfrac{4 m_q^2  N_c}{(4\pi)^2f_\pi^2}m_\pi^2F^{\prime}(m_\pi^2)
\right]\dfrac{\Delta}{m_q}
+V_{\rm fin}\;.
\label{fullb}
\eqa
The finite-temperature part of the one-loop effective potential at $\mu_B=0$ is
\bqa
V_{T}&=&-2N_cT\int_p\bigg\{
\log\Big[1+e^{-\beta E_{u}}\Big]
+\log\Big[1+e^{-\beta E_{d}}\Big]
+\log\Big[1+e^{-\beta E_{\bar{u}}}\Big]
+\log\Big[1+e^{-\beta E_{\bar{d}}}\Big]
\bigg\}\;.
\label{fd}
\eqa
\end{widetext}
The complete one-loop effective potential in the QM model in 
the large-$N_c$ limit 
is then the sum of
Eqs. (\ref{fullb}) and (\ref{fd}).
Note that Eq. (\ref{fd}) vanishes at $T=0$ and that the
only $\mu_I$-dependence of $V_{\rm1-loop}$ is line three of Eq. (\ref{fullb}).
\section{Coupling to the Polyakov loop}
In a pure gauge theory, the Polyakov loop is an order parameter
for deconfinement, as first discussed in Refs. \cite{yaffe1,yaffe2}.
In QCD with dynamical quarks, it is an approximate order parameter.
This is analogous to the quark condensate which is an exact order parameter
for chiral symmetry for
massless quark but only an approximate order parameter for massive quarks.
The Polyakov loop is defined as the trace of the thermal Wilson
line, where the thermal Wilson line $L$ is given by
\bqa
L({\bf x})&=&{\cal P}\exp\left[
i\int_0^{\beta}d\tau A_4({\bf x},\tau)
\right]\;,
\label{ldef}
\eqa
where $A_4=iA_0$ is the temporal component of the gauge field in Euclidean 
space, $A_0=t_aA_0^a$, $t_a={1\over2}\lambda^a$ are the generators of
$SU(3)_c$ gauge group, $\lambda^a$ are the Gell-Mann matrices, and 
${\cal P}$ denotes path ordering.
The background field $A_4$ in the Polyakov gauge is
\bqa
A_4&=&t_3A_4^{3}+t_8A_4^{8}\;,
\label{gauge}
\eqa
where $A_4^{3}$ and $A_4^{8}$ are time independent fields. 
Substituting Eq. (\ref{gauge}) into Eq. (\ref{ldef}), the Wilson line becomes
\bqa
L=\left(
\begin{array}{ccc}
e^{i(\phi_1+\phi_2)}&0&0\\
0&e^{i(-\phi_1+\phi_2)}&0\\
0&0&e^{-2i\phi_2}
\end{array}\right)\;,
\eqa
where we have defined $\phi_1={1\over2}\beta A_4^{3}$
and $\phi_2={1\over2\sqrt{3}}\beta A_4^{8}$.
Introducing the Polyakov loop variables\footnote{We express
the various contributions to the effective potential in terms
of $\Phi$ and $\bar{\Phi}$, although they are equal in the present case.}
\bqa
\Phi = {1 \over N_c} {\rm Tr}L \;, \quad \bar{\Phi} = {1 \over N_c} {\rm Tr}
L^\dagger\;,
\eqa
the finite-temperature fermion contribution can then be written as 
\begin{widetext}
\bqa\nonumber
V_T&=&-2T\int{d^3p\over(2\pi)^3}
\bigg\{
{\rm Tr}\log\Big[1+3(\Phi+\bar{\Phi}e^{-\beta E_u})e^{-\beta E_u}+e^{-3\beta E_u}\Big]
+{\rm Tr}\log\Big[1+3(\bar{\Phi}+\Phi e^{-\beta E_{\bar{u}}})e^{-\beta E_{\bar{u}}}
+e^{-3\beta E_{\bar{u}}}\Big]
\\
&&
+{\rm Tr}\log\Big[1+3(\Phi+\bar{\Phi}e^{-\beta E_d})e^{-\beta E_d}+e^{-3\beta E_d}\Big]
+{\rm Tr}\log\Big[1+3(\bar{\Phi}+{\Phi} e^{-\beta E_{\bar{d}}})e^{-\beta E_{\bar{d}}}
+e^{-3\beta E_{\bar{d}}}\Big]
\bigg\}\;.
\label{real}
\eqa
\end{widetext}
Eq. (\ref{real}) reduces to Eq. (\ref{fd}) upon setting 
$\Phi=\bar{\Phi}=1$, i.e. we obtain the finite-temperature part of
the effective potential in the quark-meson model.

The Polyakov loop has now been coupled to the quark sector of the model; 
we next need to include the contribution to the free energy 
density from the gauge sector. This is a phenomenological potential,
which is a function of $\Phi$ and $\bar{\Phi}$, and is required
to reproduce the pressure for pure-glue QCD calculated on the lattice
for temperatures around the transition temperature.
There are several potentials on the market \cite{ratti,ratti2,pawlow,fukushimi}
with similar properties. We will first be                              
using the polynomial potential of Ref. \cite{ratti}
\bqa
{{\cal U}\over {T^4}}&=&-{1\over2}b_2 \Phi\bar{\Phi} 
-{1\over6}b_3\left(
\Phi^3+\bar{\Phi}^3
\right)
+{1\over4}b_4\left(\Phi\bar{\Phi}\right)^2\;,
\label{pg}
\eqa
where the constants are
\bqa
b_2&=&a_0+a_1\left({T_0\over T}\right)
+a_2\left({T_0\over T}\right)^2
+a_3\left({T_0\over T}\right)^3\;,\\
b_3&=&{3\over4}\;,\\
b_3&=&{30\over4}\;,
\eqa
with $a_0=6.75$, $a_1=-1.95$, $a_2=2.625$, and, $a_3=-7.44$.\\
We will also use the logarithmic Polyakov-loop potential
of Ref.~\cite{ratti2}
\bqa\nonumber
    {{\cal U}\over {T^4}}&=&-{1\over2}a \Phi\bar{\Phi} +
    b \log\big[ 1 - 6\Phi\bar{\Phi} +4(\Phi^3 +\bar{\Phi}^3)
\\ &&
      -3(\Phi\bar{\Phi}
      )^2 \big]\;,
\label{pglog}
\eqa
with
\bqa
a &=& 3.51 -2.47\left({T_0\over T}\right) +15.2\left({T_0\over T}\right)^2\;, \\
b &=& -1.75\left({T_0\over T}\right)^3\;.
\eqa
The temperature $T_0$ is defined by
\bqa
T_0(N_f,\mu_I)&=&T_{\tau}e^{-1/(\alpha_0b(\mu_I))}\;,
\eqa
where we have modeled the $\mu_I$-dependence in the same way as the
$\mu_B$-dependence in  \cite{pawlow}
\bqa
b(\mu_I)&=&{1\over6\pi}(11N_c-2N_f)-b_{\mu_I}{\mu_I^2\over T_{\tau}^2}\;.
\label{bmu}
\eqa
The parameter $T_{\tau}=1.77\;\text{GeV}$ and $\alpha_0 = 0.304$ are determined
such that the transition temperature for pure glue at $\mu_I=0$ is $T_0=270$ MeV
\cite{t270}. 
The curvature of the deconfinement transition in $\mu_I$ direction is governed
by $b_{\mu_I}$, which is chosen as
\bqa
b_{\mu_I} = {16 \over \pi}N_f. 
\eqa
The full thermodynamic potential is now given by the sum of
Eqs. (\ref{fullb}), (\ref{real}), and (\ref{pg}) or (\ref{pglog}) respectively.
From Eqs. (\ref{eu})--(\ref{ed}), it is easy to see that
Eq. (\ref{real}) is real, thus there is no sign problem at $\mu_B=0$.
We also note that Eqs. (\ref{real}), (\ref{pg}) and (\ref{pglog})
vanish in the limit
$T\rightarrow0$ and the PQM model therefore reduces to the QM
model.

\begin{figure}[htb]
\includegraphics[width=0.45\textwidth]{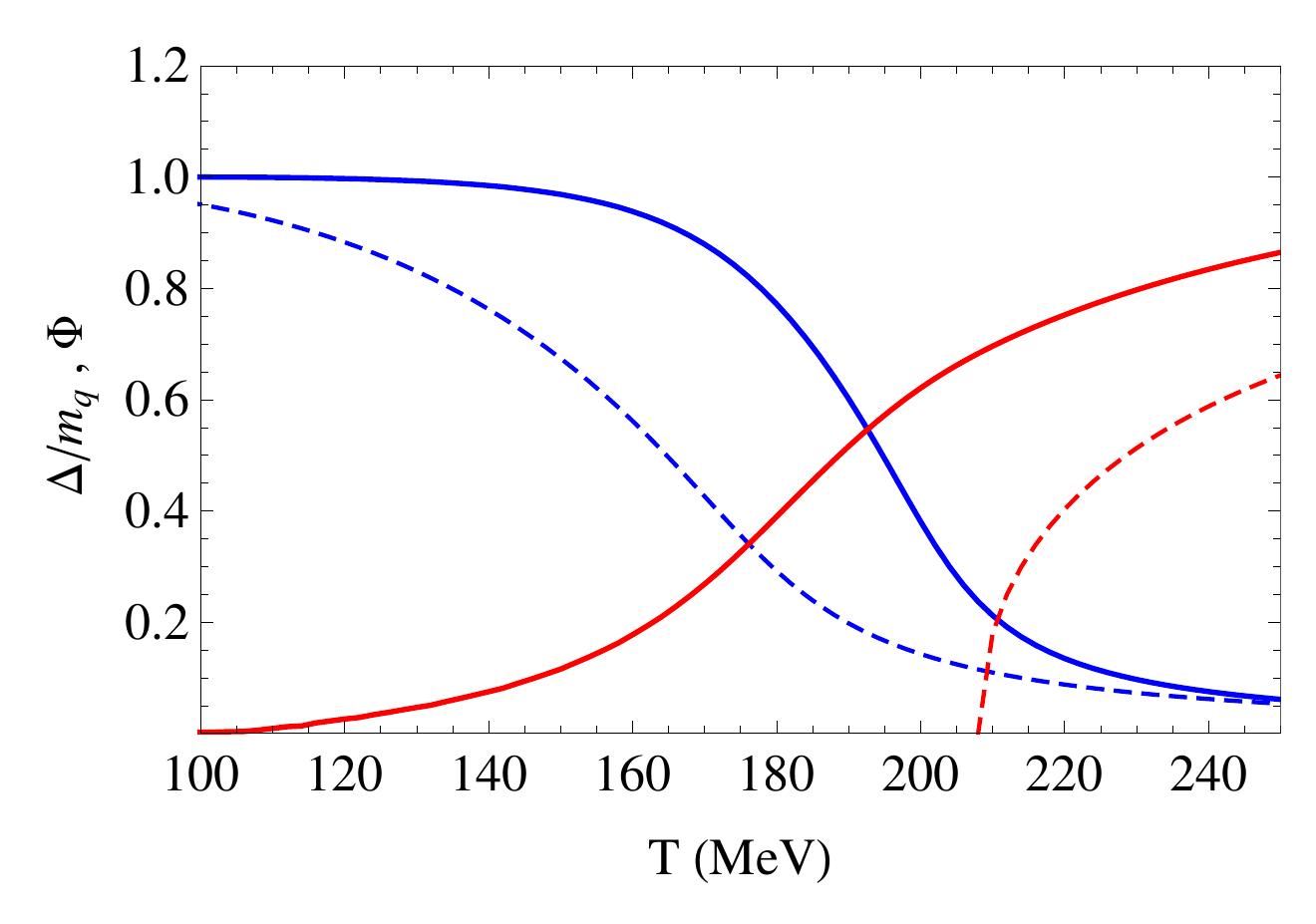}
\caption{Normalized chiral condensate ${\Delta\over m_q}$ 
(blue lines) and 
Polyakov-loop $\Phi$ (red lines) 
as functions of the temperature $T$ for $\mu_B=\mu_I=0$. See main text
for details.}
\label{mu0}
\end{figure}

In Fig. \ref{mu0}, we show the normalized chiral condensate 
${\Delta\over m_q}$ 
(blue lines) and the expectation value of the
Polyakov loop $\Phi$ as functions of the temperature
$T$ at $\mu_B=\mu_I=0$ using the polynomial potential~(\ref{pg}).
The blue dashed line
is the chiral condensate obtained in QM model while the 
blue solid line is obtained in the PQM model, i.e. with the coupling
between the order parameters. Similarly, the red dashed line is
obtained using the pure-glue potential for $\Phi$ (with the $N_f$
dependent $T_0=208\;\text{MeV}$), while the red solid line is obtained in the 
PQM
model. We notice that the critical temperature for the 
chiral transition moves to the right, i.e. to higher temperatures while 
the transition temperature for deconfinement moves to the left.
They are now within a few MeV of each other, with the deconfinement transition 
occurring at slightly lower temperature than the chiral transition.

\section{Phase diagram}
In this section, we discuss the phase diagram in the 
$\mu_I$--$T$ plane. In the numerical work below, we set
$N_c=3$, $m_{\pi}=140$ MeV, and $f_{\pi}=93$ MeV.
We vary $m_{\sigma}$ between 500 and 600 MeV.

\begin{figure}[htb]
\includegraphics[width=0.45\textwidth]{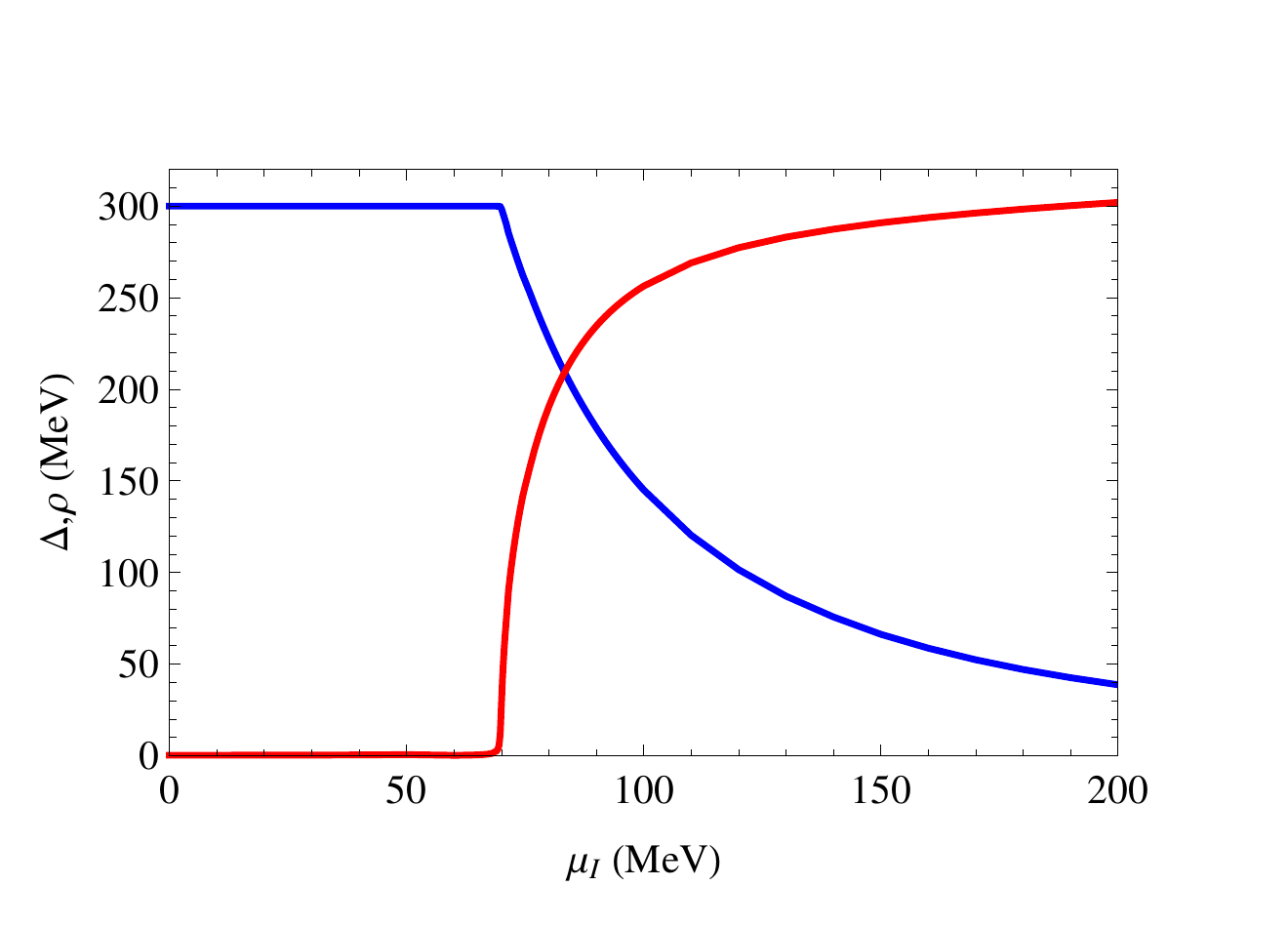}
\caption{Chiral (blue line)
and pion condensates (red line) $\Delta$ and 
$\rho$ as functions of the isospin
chemical potential $\mu_I$ at $T=0$.}
\label{t0}
\end{figure}

In Fig. \ref{t0}, we show the chiral (blue line)
and pion condensates (red line) as functions
of $\mu_I$ at zero temperature. We notice the onset of pion condensation
which takes place at exactly $\mu_I={1\over2}m_{\pi}$ 
as we will discuss in some detail below.
Moreover, the quark condensate decreases with $\mu_I$
once the pion condensate is nonzero.
Finally, all physical quantities, 
are independent of
$\mu_I$ from $\mu_I=0$ all the way up to $\mu_I={1\over2}m_{\pi}$.
For example, the effective potential is
independent of $\mu_I$, implying via Eq. (\ref{na}) that 
the isospin density vanishes.
This is an example of the Silver Blaze property \cite{cohen}
and was discussed 
in detail in the context of pion condensation in Refs. \cite{lorenz,patrick}. 
We refer to this region as the vacuum phase.

\begin{figure}[htb]
\includegraphics[width=0.45\textwidth]{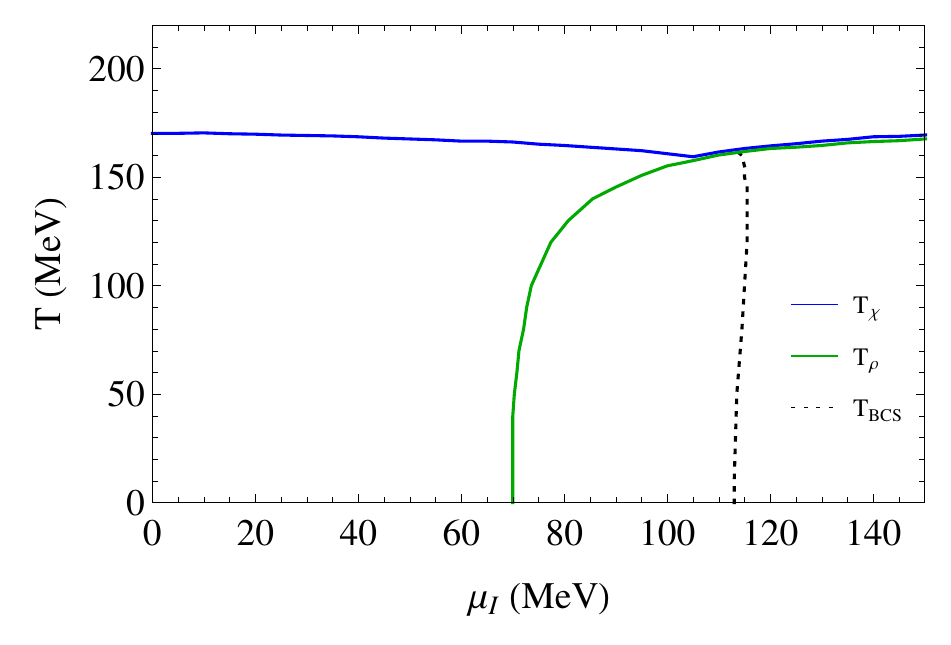}
\caption{Phase diagram in the $\mu_I$--$T$ plane for $\mu_B=0$
without Polyakov loop. See main text for details.}
\label{phase00}
\end{figure}

In Fig. \ref{phase00}, we show the phase diagram 
in the $\mu_I$--$T$ plane for $\mu_B=0$ without the Polyakov loop, i.e.
for the quark-meson model. The blue line is the transition line for the
chiral transition and the green line is the transition line for
condensation of $\pi^{+}$,
The blue line is defined by the inflection point of 
the order parameter $\Delta$ as functions of $T$
for fixed $\mu_I$.
and the black dotted line indicates the
crossover from a pion condensate to a BCS state with Cooper pairs.

The onset of pion condensation at $T=0$ is for $\mu_I={1\over2}m_{\pi}$, which
is guaranteed by the way we have determined the parameters in the
Lagrangian. This was explicitly demonstrated in Ref. \cite{patrick}. 
We can understand this result by considering the energy of a zero-momentum 
pion in the vacuum phase is $m_{\pi}-2\mu_I$. If condensations of pions
is a second order  transition,  it  must  take  place  exactly  at  a  point
where the (medium-dependent) mass of the pion drops
to zero, because in the condensed phase there is a massless 
Nambu-Goldstone mode associated with the breaking of a
$U(1)$ symmetry.
If one uses matching at tree level, there
will be finite corrections to this relation.  Likewise, if one
uses the effective potential itself to define the pion mass,
one uses the pion self-energy at zero external momentum
and so the pole of the propagator is not at the physical
mass.  Again there will be finite corrections 
and in some cases, the deviation from the exact result
can be significant \cite{lorenz}.

The condensation of pions is always a second-order
transition with mean-field critical exponents.
The order of the transition to a BEC is in agreement with the
functional renormalization group application to the QM model
in Ref. \cite{lorenz}.
The critical isospin chemical potential is fairly constant for temperatures
up to approximately $T=100$ MeV, after which it rapidly increases. For large 
values of $\mu_I$ the critical temperature for pion condensation stays at 
$T_\rho\approx 187$ MeV.
We also notice that the chiral transition temperature
line $T_{\chi}$ meets
the critical temperature line for pion condensation $T_{\rho}$ at
$\left(\mu_I^{\rm meet}, T^{\rm meet}\right) \approx (105, 159)$ MeV, and coincide
for larger values of $\mu_I$.

As we have seen, we enter the BEC phase when $\mu_I$ exceeds
${1\over2}m_{\pi}$.
As $\mu_I$ increases the quark mass $\Delta$ decreases as shown in 
Fig. \ref{t0}. Once $\mu_I>\Delta$, 
the $u$-quark and $\bar{d}$-quark energies,  
Eqs. (\ref{eu}) and (\ref{ed}), are no longer minimized for ${p}=0$, but
for $p=\sqrt{\mu_I^2-\Delta^2}$. This change is a signal of the
BEC-BCS crossover.
Although the BEC-BCS crossover is not particularly sharp, it is typically
defined by the condition $\mu_I>\Delta$ \cite{sun,tomas}.
The crossover starts at $\mu_I=113$ MeV for $T=0$ and is almost
independent
of the temperature, as can be seen from the
Figure.~


\begin{figure}[htb]
\includegraphics[width=0.45\textwidth]{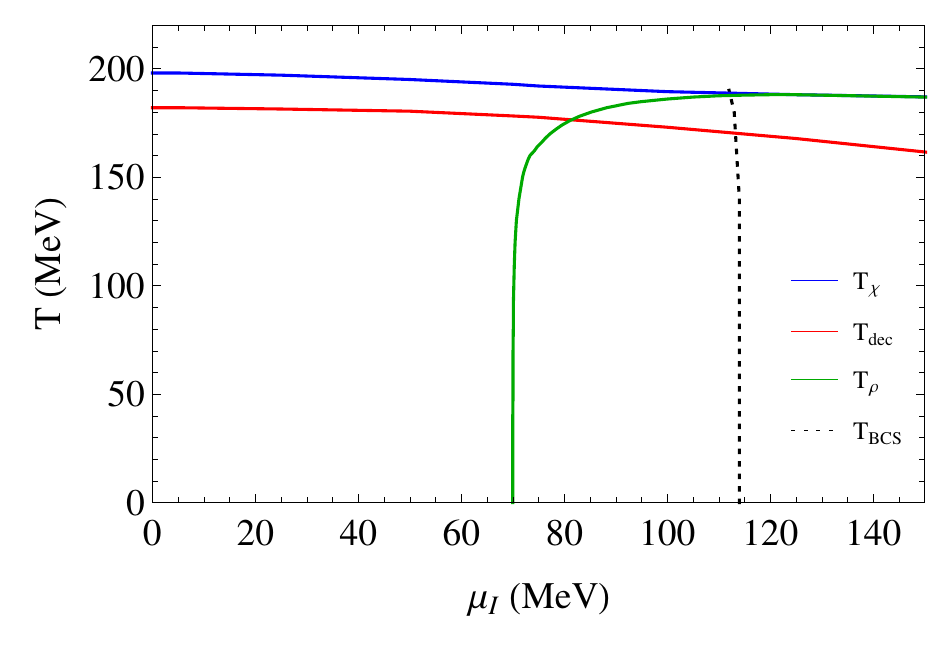}
\caption{Phase diagram in the $\mu_I-T$ plane for $\mu_B=0$ with the Polyakov-loop potential Eq. (\ref{pg}). See main text for details.}
\label{phase0}
\end{figure}

In Fig. \ref{phase0}, we show the phase diagram  in the $\mu_I$--$T$ plane
at zero baryon chemical potential with the Polyakov loop
and ${{\cal U}\over T^4}$ given by (\ref{pg}). 
The green line is the critical line for Bose-Einstein condensation of
charged pions, the red line is  the critical line for deconfinement,
and the blue line is the critical line for the chiral transition.
Finally, the black dotted line indicates the BEC-BCS transition line.
The blue and red lines are defined by the inflection point of 
the order parameters $\Delta$ and $\Phi$ as functions of $T$
for fixed $\mu_I$.
As in the QM model, the transition temperature line $T_{\chi}$ joins
the critical temperature for pion condensation at
$\left(\mu_I^{\rm meet}, T^{\rm meet}\right) \approx (116, 187)$ MeV.
The transition line for deconfinement lies approximately 15 MeV below the
chiral transition line for $\mu_I=0$ increasing somewhat
for large values of $\mu_I$.

\begin{figure}[htb]
\includegraphics[width=0.45\textwidth]{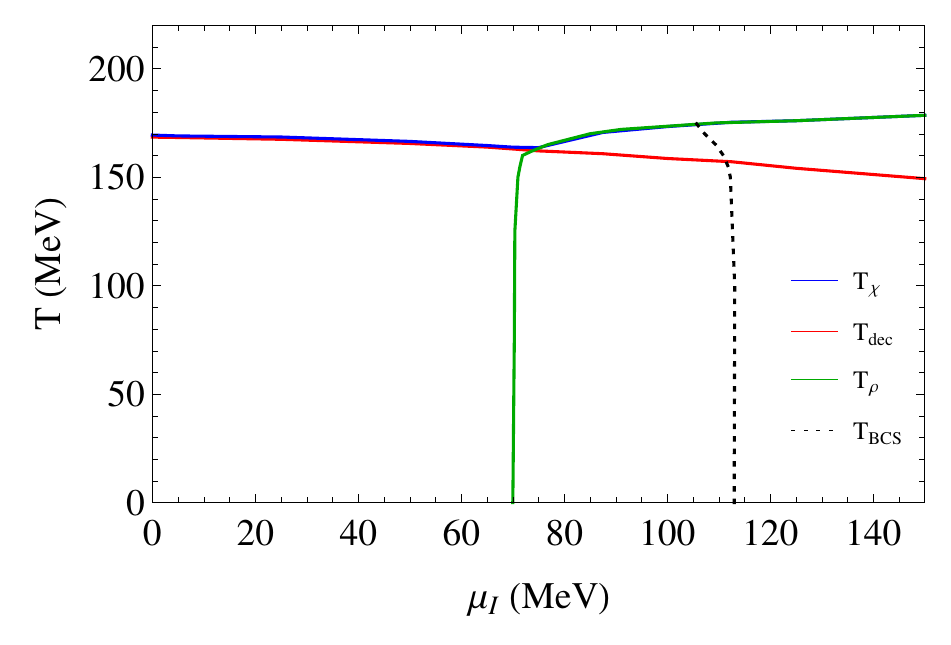}
\caption{Phase diagram in the $\mu_I-T$ plane for $\mu_B=0$ with the Polyakov-loop potential Eq. (\ref{pglog}). See main text for details.}
\label{phaselog}
\end{figure}

The gap between the chiral and deconfinement line can be reduced by using a
logarithmic Polyakov potential (\ref{pglog})
instead of Eq. (\ref{pg}) and decreasing the
sigma mass. For $m_\sigma = 500$ MeV the two lines basically coincide at $T=0$
as seen in Fig. \ref{phaselog}. 
The chiral and deconfinement transition line
also meet the pion-condensed line at a point 
for a smaller value of $\mu_I$ as compared to Fig.~\ref{phase0},
$\left(\mu_I^{\rm meet}, T^{\rm meet}\right) \approx (75, 164)$ MeV.

\begin{figure}[htb]
\includegraphics[width=0.45\textwidth]{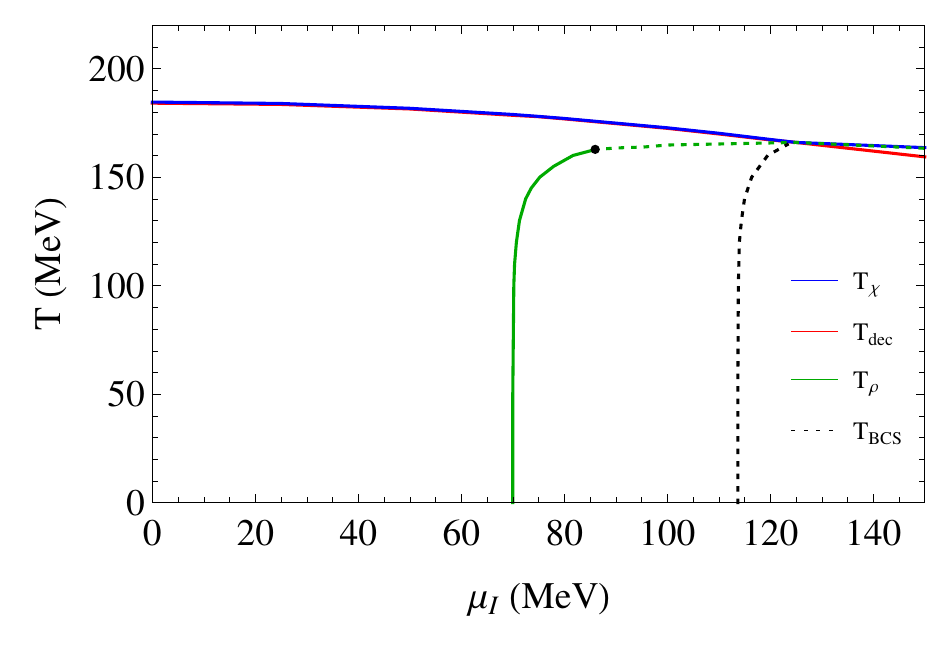}
\caption{Phase diagram in the $\mu_I-T$ plane for $\mu_B=0$ in the standard mean field approximation with the Polyakov-loop potential Eq. (\ref{pg}). See main text for details.}
\label{phasesmfa}
\end{figure}

For completeness, we show in Fig. \ref{phasesmfa} the phase diagram in the
standard mean-field approximation (sMFA), which is a common approximation used
in the literature, where one ignores 
the loop corrections to the vacuum potential, i.e. uses Eq.~(\ref{treepot})
instead of Eq.~(\ref{fullb}).
We find the critical temperature for pion condensation to be smaller than
for the one-loop potential in Fig. \ref{phase0}. We also find a first-order
transition of the pion condensate above a critical isospin chemical potential
$\mu_I \approx 86$ MeV, indicated by the black dot in the figure. This critical
point is absent once we go beyond the sMFA, at least in the region of $\mu_I$
considered here.


Our phase diagram is in qualitative good agreement with that obtained
by Brandt, Endr\H{o}di, Schmalzbauer using lattice 
simulations \cite{gergy1,gergy2,gergy3}, in particular if we use
a logarithmic Polyakov loop potential and choose a lower sigma mass of
$500$ MeV.
We believe that the quantitative
differences 
(essentially the temperature dependence of the various transition
lines)
can mainly be attributed to the fact that we have two
light flavors, while they consider 2+1 flavors; for example
the deconfinement transition temperature decreases with the number
of quarks and our transition line is consistently 
higher.\footnote{By using a smaller value of $T_0$, we can bring down the
transition line.} They find that chiral and BEC transition lines
meet at a triple point, beyond which they coincide. The latter
transition is again found to be second order for all values of $\mu_I$
and the scaling analysis is consistent with the $O(2)$
universality class.
They computed contour lines of the expectation values of 
the renormalized Polyakov loop $\Phi$ for values
0.2, 0.4, 0.6, 0.8, and 1.0. 
Given their renormalization prescription for the Polyakov 
loop, developed in  \cite{renpol}, a
possible choice for $T_{\rm dec}$ is $\Phi=1$, which implies
that it coincides with $T_{\chi}$ within errors \cite{gergy3}.
Finally, we mention that the deconfinement line
penetrates smoothly into the BEC phase and that they identify
this line with the BEC-BCS transition.



\section*{Acknowledgments}
The authors would like to thank G. Endr\H{o}di for useful discussions.

\appendix
\section{Integrals}
With dimensional regularization,
the momentum integral is generalized to $d=3-2\epsilon$
spatial dimensions. We define the dimensionally regularized integral by
\bqa
\int_p&=&
\left({e^{\gamma_E}\Lambda^2\over4\pi}\right)^{\epsilon}
\int{d^{d}p\over(2\pi)^d}\;,
\label{dint}
\eqa
where $\Lambda$ is the renormalization scale in the
modified minimal subtraction scheme $\overline{\rm MS}$.
\begin{widetext}
In order to calculate the vacuum part of the 
effective potential, we need the vacuum integrals
\bqa
\int_p\sqrt{p^2+M^2}
&=&-{M^4\over(4\pi)^2
}\left({e^{\gamma_E}\Lambda^2\over M^2}\right)^{\epsilon}
\Gamma(-2+\epsilon)
=-{M^4\over2(4\pi)^2
}\left({\Lambda^2\over M^2}\right)^{\epsilon}
\left[{1\over\epsilon}+{3\over2}+{\cal O}(\epsilon)\right]\;,
\label{i1}
\\ 
\int_p{1\over(p^2+M^2)^{3\over2}}
&=&{4\over(4\pi)^2}
\left({e^{\gamma_E}\Lambda^2\over M^2}\right)^{\epsilon}
\Gamma(\epsilon)
={4\over(4\pi)^2}
\left({\Lambda^2\over M^2}\right)^{\epsilon}
\left[{1\over\epsilon}+{\cal O}(\epsilon)\right]\;.
\label{i3}
\eqa
\section{Parameter fixing}
In this Appendix, we briefly discuss the fixing of the model parameters.
At tree level, the relations between these parameters and the
physical quantities are given by Eqs. (\ref{rel1})--(\ref{rel2}).
In the
on-shell scheme, the counterterms are chosen such that they exactly
cancel the loop corrections to the self-energies 
and couplings evaluated on the mass shell, 
and such that the residues evaluated on shell are unity.
Consequently, the renormalized parameters are independent
of the renormalization scale and satisfy the tree-level 
relations \cite{sir1,sir2,hollik}.
In the $\overline{\rm MS}$ scheme, the counterterms are chosen
so that they cancel only the poles in $\epsilon$ of the loop corrections.
The bare parameters
are the same in the two schemes, which means that we can relate
the corresponding renormalized parameters.
The running parameters in the $\overline{\rm MS}$ scheme
can therefore be expressed in terms of the 
physical masses $m_{\sigma}$, $m_{\pi}$, and $m_q$ as well as the pion decay constant $f_\pi$.
In Ref. \cite{crew} we found 
\bqa \nonumber
m^2_{\ms} 
&=& m^2 +8ig^2N_c \left[ A(m_q^2) +\mbox{$1\over4$}(m_\sigma^2-4m_q^2)
B(m_\sigma^2) 
-\mbox{$3\over4$}m_{\pi}^2B(m_{\pi}^2)\right] -\delta m^2_{\ms} \\
 &=& m^2+
\dfrac{4g^2N_c}{(4\pi)^2} 
\left[m^2\log\mbox{$\Lambda^2\over m_q^2$}
-2m_q^2-{1\over2}\left(m_\sigma^2-4m_q^2\right)F(m_\sigma^2) 
+{3\over2}{m_\pi^2}F(m_\pi^2) \right]\;, 
\label{osm1}
\\ \nonumber
\lambda_{\ms} 
&=& \lambda -\dfrac{12ig^2N_c}{f_{\pi}^2} (m_\sigma^2-4m_q^2)B(m_\sigma^2) 
+\dfrac{12ig^2N_c}{f_{\pi}^2}m_\pi^2B(m_\pi^2) 
-4i\lambda g^2N_c\left[B(m_{\pi}^2)+m_{\pi}^2B^{\prime}(m_{\pi}^2)\right]
- \delta\lambda_{\ms}
\\ \nonumber &=& 
\lambda
+\bigg\{\dfrac{12g^2N_c}{(4\pi)^2f_\pi^2}\left[(m_\sigma^2-4m_q^2)\left(
\log\mbox{$\Lambda^2\over m_q^2$} +F(m_\sigma^2)\right)
+m_\sigma^2\left(
\log\mbox{$\Lambda^2\over m_q^2$}+F(m_\pi^2)+m_\pi^2F^{\prime}(m_\pi^2)\right)
\right.\\ &&\left.
-m_\pi^2\left(
2\log\mbox{$\Lambda^2\over m_q^2$}+2F(m_\pi^2)
+F^{\prime}(m_\pi^2)\right)\right]\bigg\}\;,
\label{osl}
\\ 
g_{\ms}^2  &= &g^2-4ig^4N_c
\left[B(m_\pi^2) +m_\pi^2B^\prime(m_\pi^2) \right] -\delta g_{\ms}^2 
={m_q^2\over f_{\pi}^2}\left\{1 + \dfrac{4g^2N_c}{(4\pi)^2}\left[
\log\mbox{$\Lambda^2\over m_q^2$}+F(m_\pi^2) +m_\pi^2F^\prime(m_\pi^2)
\right]\right\}\;,
\label{g00}
\\ 
h_{\ms}&=&h-2ig^2N_cm_\pi^2f_\pi\left[
B(m_\pi^2)-m_\pi^2B^{\prime}(m_\pi^2)\right]-\delta h_{\ms}
=h
\left\{
1+{2g^2N_c\over(4\pi)^2}
\left[\log\mbox{$\Lambda^2\over m_q^2$}
+F(m_\pi^2)-m_\pi^2F^{\prime}(m_\pi^2)\right]\right\}\;,
\label{hhh}
\eqa
\end{widetext}
where $A(m_q^2)$, $B(p^2)$, and $B^{\prime}(p^2)$
are integrals in $d=4-2\epsilon$ dimensions in Minkowski space. 
Going to Euclidean space, they can be straightforwardly computed and
read
\bqa\nonumber
A(m^2_q)&=&\int_k{1\over k^2-m^2_q}
\\&=&{im^2_q\over(4\pi)^2}\left({\Lambda^2\over m_q^2}\right)
\left[{1\over\epsilon}+1+{\cal O}(\epsilon)\right]\;,
\label{adef}
\\ \nonumber
B(p^2)&=&
\int_k{1\over(k^2-m_q^2)[(k+p)^2-m_q^2]}
\\ &=&
{i\over(4\pi)^2}\left({\Lambda^2\over m_q^2}\right)
\left[{1\over\epsilon}+F(p^2)
+{\cal O}(\epsilon)\right]\;,
\label{bdef}
\\
B^{\prime}(p^2)&=&{i\over(4\pi)^2}F^{\prime}(p^2)\;,
\eqa
where we have defined 
\bqa
F(p^2)&=&2-2r\arctan\left({1\over r}\right)\;,
\\
F^{\prime}(p^2)&=&{4m_q^2r\over p^2(4m_q^2-r^2)}\arctan\left({1\over r}\right)
-{1\over p^2}\;,
\eqa
with $r=\sqrt{{4m_q^2\over p^2}-1}$.


The running parameters satisfy 
the following renormalization group equations
\bqa
\label{run1}
\Lambda{dm_{\ms}^2(\Lambda)\over d\Lambda}&=&{8 N_c m^2_{\ms}(\Lambda)g_{\ms}^2
(\Lambda)
\over(4\pi)^2}\;,
\\
\Lambda{dg_{\ms}^2(\Lambda)\over d\Lambda}&=&{8 N_c g_{\ms}^4(\Lambda)
\over(4\pi)^2}
\;,
\\
\Lambda{d\lambda_{\ms}(\Lambda)\over d\Lambda}&=&{16N_c\over(4\pi)^2}
\left[\lambda_{\ms}(\Lambda) g^2_{\ms}(\Lambda)-6g^4_{\ms}(\Lambda)
\right]\;,
\\
\Lambda{dh_{\ms}(\Lambda)\over d\Lambda}&=&{4 N_c g^2_{\ms}(\Lambda)
h_{\ms}(\Lambda)\over(4\pi)^2}
\label{run4}
\;.
\eqa
The solutions to Eqs. (\ref{run1})--(\ref{run4}) are
\bqa
\label{sol1}
m_{\ms}^2(\Lambda)&=&
{m_0^2\over1-{4g_0^2N_c\over(4\pi)^2}
\log{\Lambda^2\over \Lambda_0^2}
}\;.
\\
g_{\ms}^2(\Lambda)&=&
{g_0^2\over1-{4g_0^2N_c\over(4\pi)^2}
\log{\Lambda^2\over \Lambda_0^2}
}\;,
\\
\lambda_{\ms}(\Lambda)&=&{\lambda_0-{48g_0^4N_c\over(4\pi)^2}
\log{\Lambda^2\over \Lambda_0^2}
\over\left(1-{4g_0^2N_c\over(4\pi)^2}
\log{\Lambda^2\over \Lambda_0^2}
\right)^2}\;,
\label{sol3}
\\
h_{\ms}(\Lambda)&=&
{h_0\over1-{2g_0^2N_c\over(4\pi)^2}
\log{\Lambda^2\over \Lambda_0^2}
}\;,
\label{sol5}
\eqa
where $m_0^2$, $g_0^2$, $\lambda_0$ and $h_0$, are the 
values of the running parameters at the scale $\Lambda_0$.
We choose $\Lambda_0$ to satisfy
\bqa
\log{\Lambda_0^2\over m_q^2}+F(m_\pi^2)+m_\pi^2F^{\prime}(m_\pi^2)
&=&0\;.
\label{l0}
\eqa
One can now evaluate Eqs. (\ref{osm1})--(\ref{hhh}) at the scale
$\Lambda=\Lambda_0$ 
to find
$m_0^2$, $\lambda_0$, $g_0^2$, and $h_0$. Inserting Eqs. 
(\ref{sol1})--(\ref{sol5}) into Eq. (\ref{veff})
using the results for $m_0^2$, $\lambda_0$, $g_0^2$, and $h_0$, 
we obtain the final result Eq. (\ref{fullb}).


\bibliography{refs}{}
\bibliographystyle{apsrmp4-1}
\end{document}